\documentclass[smallcondensed]{svjour3}
\usepackage{amsmath}
\usepackage{multirow}
\usepackage[usenames, dvipsnames]{color}
\usepackage{colortbl}

\usepackage{balance}
\usepackage{array}
\usepackage{times}
\usepackage{tabularx}

\usepackage{picture}
\usepackage{wrapfig}
\usepackage{algorithm}
\usepackage{graphicx}
\usepackage{algorithmicx}
\usepackage{algpseudocode}

\newcommand{\bi}{\begin{itemize}}
\newcommand{\ei}{\end{itemize}}
\newcommand{\be}{\begin{enumerate}}
\newcommand{\ee}{\end{enumerate}}
\newcommand{\tion}[1]{\S\ref{sect:#1}}
\newcommand{\fig}[1]{Figure~\ref{fig:#1}}

\usepackage{cite}
\newcommand{\subparagraph}{}
\usepackage{url}

\usepackage{colortbl}
\DeclareMathSizes{7}{7}{7}{7} 
\titlerunning{Are Delayed Issues Harder to Resolve? } 

\begin{document}
\date{}
 \title{Are Delayed Issues Harder to Resolve? Revisiting Cost-to-Fix of Defects throughout the Lifecycle}
 
 \author{Tim Menzies, William Nichols, Forrest Shull, Lucas Layman}
 
\institute{
Tim Menzies \at
       CS, North Carolina State University, USA,
                \email{tim.menzies@gmail.com} \
\and 
William Nichols, Forrest Shull \at 
        Software Engineering Institute ,
        Carnegie Mellon University, USA.
        \email{\{wrn,fjshull\}@sei.cmu.edu} 
\and  
Lucas Layman \at
        Fraunhofer CESE,  
        College Park, USA,
       \email{llayman@cese.fraunhofer.org}
}
\maketitle
\begin{abstract}
Many  practitioners and academics
believe in a delayed issue effect (DIE); i.e.
 the longer an issue lingers in the system, the more effort it requires to resolve.
This belief
is often  used to justify 
major investments in  new development
processes that promise to retire more issues sooner.

This paper tests for the delayed issue effect in
171 software projects conducted around the world in the period from 2006--2014.
To the best of our knowledge,  this is the largest study
yet published on this effect.
We found no evidence for the  delayed issue effect; i.e.
the  effort  to resolve 
issues in a later phase was not consistently or substantially greater than  
when  issues were resolved soon after their introduction. 

This paper documents the above study and explores reasons for this  mismatch between this common rule of thumb and empirical data.
In  summary, DIE is not some constant across all projects. Rather, DIE might
be an historical relic that  occurs intermittently 
only in  certain kinds of projects.  This is a significant result since it predicts that  new development
processes that promise to faster retire more issues will not have a guaranteed return on investment (depending on the context where applied), and that a long-held truth in software engineering should not be considered a global truism. 
\end{abstract}

 \vspace{1mm}
\noindent
{\bf Categories/Subject Descriptors:} 
D.2.8 [Software Engineering]: Process metrics.

\vspace{1mm}
  
\section{Introduction}
In 2013-2014, 
eleven  million programmers~\cite{avram14} and
half a trillion dollars~\cite{pettey14} were spent on information technology.
Such a large and growing effort should be managed and optimized via  well-researched conclusions.  
To assist in achieving this, there has been a growing recognition within the software engineering research community 
of the importance of theory building ~\cite{Sjoberg08, Paivarinta15, Stol15}. A good theory allows empirical research to go beyond simply
reporting observations and instead provides explanations for why results are observed ~\cite{Stol15}. This occurs by testing theories against data from multiple sources; by reconciling similarities and differences in results it can be determined what factors need to be accounted for in a theory ~\cite{Shull08}. Theory-building needs to be an iterative process,
in which results from practice are used to refine theories and theories are used to inform future observation and data collection ~\cite{Paivarinta15, Stol15}. It is no coincidence that it is standard practice
in other fields, such as medicine,
to continually revisit old conclusions in the light of new theories~\cite{prasad13}.

\begin{figure}[!b]  
\begin{center}
\includegraphics[,width=4in]{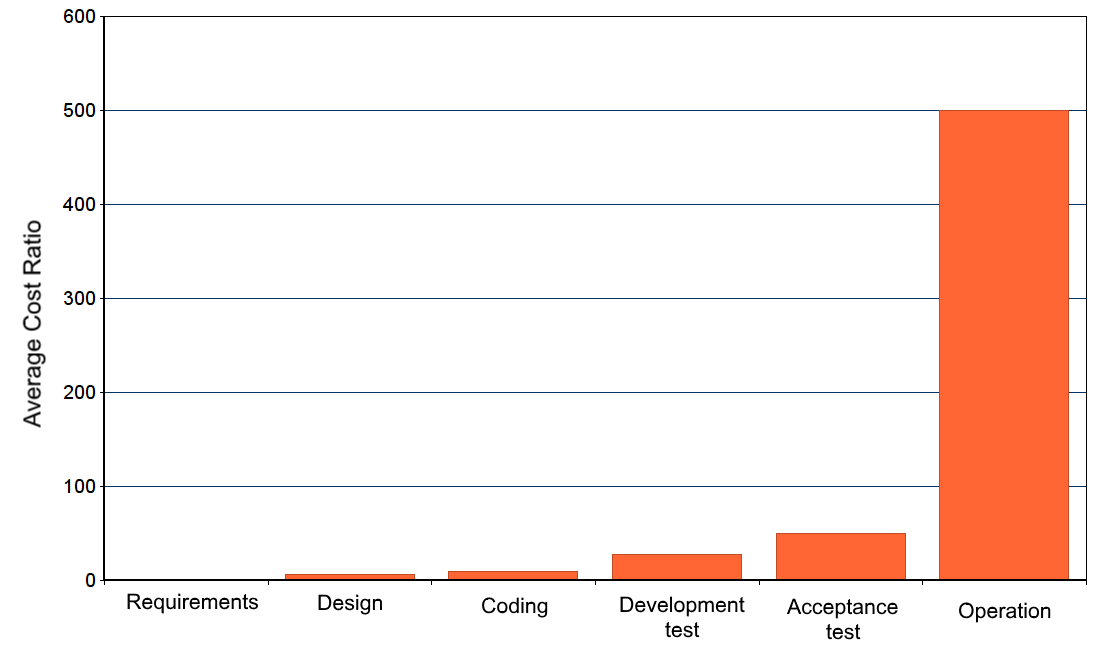}
 \end{center}
 \caption{A widely-recreated chart of the DIE effect. Adapted from Boehm'81~\cite{Boehm81}. }\label{fig:b81}
 \end{figure}

Accordingly, this paper revisits
the commonly held theory we label the {\em delayed issue effect} (hereafter, DIE):
more effort is required to resolve an issue the longer an issue lingers in a system.
 \fig{b81} shows an example of the delayed issue effect (relating
 the relative cost of fixing requirements issues at different phases of a project). As a falsifiable theory, the DIE can be compared to empirical data and, if inconsistencies are observed, refinements to the theory may be generated that better describe the phenomenon under observation ~\cite{Popper}.
 
The DIE theory is worth examination since it has been used as the basis for decision-making in software engineering. For example,  Basili and Boehm comment that, since the 1980s,  this effect
 \begin{quote}
``...has been a major driver in focusing
industrial software practice on thorough
requirements analysis and design,
on early verification and validation, and
on up-front prototyping and simulation
to avoid costly downstream fixes''~\cite{boehm01}.
\end{quote}

Like any good theory, DIE includes a rationale for why the expected results would be seen. McConnell mentions it as a ``common observation'' in the field and  summarizes the intuitive argument for why it should be so: 
\begin{quote}
''A small mistake in upstream work can affect large amounts of downstream work. A change to a single sentence in a requirements specification can imply changes in hundreds of lines of code spread across numerous classes or modules, dozens of test cases, and numerous pages of end-user documentation''~\cite{mcconnell01}. 
\end{quote}
Glass also endorses this rationale, asserting that ``requirements errors are the most expensive to fix when found during production but the cheapest to fix early in development'' is ``really just common sense''~\cite{glass02}.  Other researchers
are just as adamant in asserting that the delayed issue effect is a generally useful law of software engineering.
For example, what we call the delayed issued effect was listed at \#1 by Boehm and Basili in their ``Top 10 list'' of ``objective and quantitative data, relationships,
and predictive models that help
software developers avoid predictable pitfalls
and improve their ability to predict
and control efficient software projects''~\cite{boehm01}.
  
In analyzing data from a contemporary set of software development projects, however, we did not find results to corroborate these claims. While the delayed issue effect might have been a dominant
effect decades ago, this does not mean that it is necessarily so for $21^{\mathit{st}}$ century
software development. 
 The delayed issue effect was first reported in 1976 in a era of punch card programming
and non-interactive environments~\cite{Boehm76}. In the 21$^\mathit{st}$ century, we  program in 
interactive environments with higher-level languages and better source code control
tools. Such tools allow for the faster refactoring of existing
code-- in which case, 
managing the changes required to fix (say) an incorrect requirements assumption
is far less    onerous    than before. Further, software engineering theory and practice has evolved into new paradigms focused on rapid feedback and delivery, enabled by significant technological advances in the past 40 years. There is little empirical evidence for the delayed issue effect since its initial observation, no doubt due in part to DIE being ``just common sense'' as Glass states~\cite{glass02}. 

This article explores the currency of the delayed issue effect.
After some initial definitions, we discuss the
value of checking old ideas. Next, we present a survey of industrial practitioners and researchers that documents the widespread belief that delayed issues have a negative impact on projects.  After that, we  analyze 171 software  projects developed in the period 2006--2014 and find  {\em no evidence} of the delayed issue effect. Finally, we discuss the validity and implications of our results, as well as possible reasons for the lack of observed effect given the state of the practice - reasons which, when subjected to further testing, may prove useful for refining the theory. To ensure reproducibility,
all the data  used in this study is available in the PROMISE
repository at openscience.us/repo. 
To the best of our knowledge,
this the largest study devoted the delayed issue effect yet conducted.

\subsection{Preliminaries}\label{sect:nontsp}
Before beginning, it is appropriate to make the following full disclosure statement. 
All 171 software projects
studied here were developed using the Team Software Process (TSP$\textsuperscript{SM}$), which is a software development methodology
developed and promulgated by the employer of the second and third author of this paper (for more details on TSP,
 see \tion{tsp}).

We argue that TSP is not such a radical change to software development that it can
stamp out a supposedly rampant problem like the delayed issue effect. We view TSP as a better way to
 {\em monitor} the activities   of  existing projects.  TSP
 does not significantly change a project-- it just offers a better way to log the activity within
 that project. The limitations of our sample drawing from TSP projects are discussed more thoroughly in the Threats to Validity section. 
 

\section{Definitions \& Claims}
\label{sect:claims}

This paper uses the following definitions:
\bi
\item
The {\em delayed issue effect}:   it is {\em very much}  more {\em difficult} to resolve  issues in a software project, the {\em longer} they remain.
\item
 {\em Longer} time is defined as per  Boehm'81~\cite{Boehm81}; i.e. the gap between the   phases where   issues are introduced and resolved.
\item
We say that a measure $m$ collected in phase ${1,.,i,..j}$ is 
{\em very much} more when  that
   measure at phase $j$   
   is larger than the sum of those measures in the earlier phases;
   i.e. $\sum_{i=1}^{j-1} m_i $. 
\item
Issues are more {\em difficult}  
when their resolution takes more time or costs more  (e.g. needs expensive
debugging tools or the skills of expensive developers).
\ei

Note that this  definition of ``difficult to resolve''  combines two concepts: time to change and cost to change.  Is it valid to assume the equivalence of time and cost?
Certainly, there are cases where time is not the same as cost. Consider, for example, if debugging required some very expensive tool or the services or a very senior (and hence, very expensive) developer. Under those circumstances, time does not equate to cost.
Having documented the above issues, we assert that they are unlikely to be major issues in the study. One of us (Nichols) was closely associated with many of the projects in our sample. He is unaware of any frequent use of exorbitantly expensive tools or people on these projects. For more on the validity of this definition
of ``difficult to resolve'' see \tion{construct}.

This paper defends the following claim and hypothesis. The hypothesis is defended
using some statistical
significance tests while the claim is supported via a variety
of arguments.

{\bf  Claim: ``DIE'' is a  commonly held, yet poorly documented belief.}
We examine the literature promoting the DIE and find that most reference a few primary sources. Many of the papers reporting the DIE
 either (1)~are quite old (papers dating from last century);
(2)~quote prior papers without presenting   new data; 
(3)~or cite data sources that can no longer be
confirmed. We follow-up with a short survey that finds that DIE appears as the most strongly-held belief among software engineers in our sample.

{\bf Hypothesis: Delayed issues are not harder to resolve.}
 In our sample of 
 171 commercial software  projects, we offer  a statistical analysis showing that, in overwhelming majority
 of our results, there is no   
 significant increase in the time to resolve issues  as they are delayed across multiple phases.

\section{ Reassessing Old Truisms}
General
theories of software
engineering principles are common to both research and practice, although not always explicitly stated. Such theories underlie  lists of proposed general ``best practices'' for effective software development, such as
the IEEE 1012 standard for software verification~\cite{1012}. 
 Endres \& Rombach offer empirical observations, theories, and laws\footnote{Endres \& Rombach note that these are not laws of nature in the scientific sense, but theories with repeated empirical evidence.}~\cite{endres03}.
 Many other 
commonly cited researchers  do the same, e.g.,
Glass~\cite{glass02}, Jones \cite{jones07}, and Boehm~\cite{boehm00b}.
Budgen \& Kitchenham seek to reorganize SE research using
general
conclusions drawn from a larger number of studies~\cite{kitch04,budgen09}.

In contrast, there are many empirical  findings 
that demonstrate the difficulty in finding general truisms in software engineering, even for claims that seem intuitive:

\be
\item
Turhan~\cite{me12d} lists 28 studies with contradictory conclusions
on the relation of object-oriented (OO) measures to defects.  Those results
 directly  contradict some of the laws listed by 
Endres \& Rombach~\cite{endres03}.
\item
Ray et al.~\cite{ray2014lang} tested if   strongly typed languages
predict for better code quality. In  728 projects,
they found  only a modest benefit in strong typing and warn that the effect may be due to other conflating factors.
\item
Fenton \& Neil~\cite{fenton00,fenton00b}   critique the truism that
``pre-release fault rates for software
are a predictor for post-release failures'' (as claimed in~\cite{dunsmore88},
amongst others). For the systems described in~\cite{fenton97}, they
show that software modules that were highly fault-prone
prior to release revealed very few faults after release.
\item
Numerous recent {\em local learning} results compare single models
learned from all available data to multiple models learned from clusters within the data~\cite{betten14,yang11,yang13,minku13,me12d,me11m,betta12,posnett11}.
A repeated result in those studies is that the local models generated the better effort
and defect predictions (better median results,
lower variance in the predictions).
\ee
 
The dilemma of updating truths in the face of new evidence is not particular to software engineering. 
The medical profession applies  many practices based on studies
that have been disproved. For example,
a  recent article
in the Mayo Clinic Proceedings~\cite{prasad13} found  
146 medical practices based on studies 
in year $i$, but which were  reversed by subsequent trials within years $i+10$.
Even when the evidence for or against a treatment or intervention is clear, medical providers and patients may not accept it~\cite{aschwanden10}.
Aschwanden warns that ``cognitive biases''  such as  confirmation bias (the tendency to look for evidence that supports what you already know and to ignore the rest)  influence how we process information~\cite{aschwanden15}.

The cognitive issues that complicate medicine are also found in software engineering.
Passos et al.~\cite{passos11} warn that developers
usually develop their own theories of what works and what doesn't work in creating software, based on experiences from a few past
projects. Too often, these theories are assumed to be general truisms with widespread applicability to future projects. They comment ``past experiences were taken into account without 
much consideration for their context''~\cite{passos11}.
The results of J{\o}rgensen \& Gruschke~\cite{jorgensen09} support Passos et al. In an empirical study of expert effort estimation, they report that the experts rarely use lessons
  from past projects to improve their future reasoning in effort estimation~\cite{jorgensen09}. 
 They note that,
when the experts
  fail to revise their beliefs, this leads to poor
 conclusions and software projects  (see examples in~\cite{jorgensen09}).
 A similar effect is reported by
Devanbu et al.~\cite{prem16}  who examined responses from 564 Microsoft software developers from around
the world; they found that  ``(a)~programmers do indeed have very
strong beliefs on certain topics; (b)~their beliefs are primarily formed
based on personal experience, rather than on findings in empirical
research; (c)~beliefs can vary with each project, but do not necessarily
correspond with actual evidence in that project.''
Devanbu et al. further  comment that ``programmers give personal experience
as the strongest influence in forming their opinions.'' This is a troubling
result, especially given the above comments from Passos and  J{\o}rgensen et al.~\cite{passos11,jorgensen09} about how quickly practitioners form, freeze, and rarely revisit those opinions.

From all we above we conclude that, just as in medicine, 
 it is important for our field
 to regularly  reassess old truisms  like the  delayed issue effect.

\section{Motivation: ``DIE'' is commonly held, yet poorly documented}
\label{sect:belief}
One reason that industrial practitioners and academics believe so strongly in the delayed issue effect is that it is often referenced
in the SE literature. 
Yet when we look at the literature, the evidence for
delayed issue effect is both very sparse and very old.
As  shown in this section,
a goal of agile methods 
is to reduce the difficulty associated with making changes later in the lifecycle~\cite{beck00}. Yet, as shown below,
relatively little empirical data exists on this point.

We examined the literature on the delayed issue effect through a combination of snowball sampling~\cite{wohlin2014guidelines} and database search. We searched Google Scholar for terms such as ``cost to fix'' and ``defect cost'' and ``software quality cost''. The majority of the search results discuss quality measurements, quality improvement, or the cost savings of phase-specific quality improvement efforts (e.g., heuristic test case selection vs. smoke testing). A systematic literature review of software quality cost research can be found in \cite{karg2011systematic}. Relatively few articles discuss cost-to-fix as a function of when the defect was injected or found. We also conducted a general Google search for the above terms. We found a number of website articles and blog postings on this topic, e.g., \cite{IfSQ2013,Soni2016,Parker2013,Gordon2016}. From these, we gathered additional citations for the delayed issue effect, the vast majority of which were secondary sources, e.g.,~\cite{Leffingwell96,mead2004software,mcconnell1996software,mcconnell01,Tassey2002,boehm2012}. Our literature search is not exhaustive, but our results yielded an obvious trend: nearly every citation to the delayed issue effect could be traced to the seminal \textit{Software Engineering Economics}~\cite{Boehm81} or its related works~\cite{boehm88,boehm01}.\footnote{For example, popular sources such as~\cite{pressman2005software, boehm01,glass02,endres03}, with a combined citation count of over 14,500 on Google Scholar, can all trace their evidence to \textit{Software Engineering Economics}~\cite{Boehm81}.}

Ultimately, we identified nine sources of evidence for the delayed issue effect based on real project data: the original four \cite{Fagan76,Boehm76,Daly77,Stephenson76} reported in \textit{Software Engineering Economics}\cite{Boehm81}, a 1995 report by Baziuk~\cite{baziuk1995bnr} on repair costs at Nortel, a 1998 report by Willis et al.~\cite{willis1998hughes} on software projects at Hughes Aircraft, a 2002 experiment by Westland~\cite{westland2002cost} to fit regression lines to cost-to-fix of localization errors, a 2004 report by Stecklein et al.~\cite{steck04} on cost-to-fix in five NASA projects, and a 2007 survey by Reifer on CMMI Level 5 organization\cite{reifer2007profiles}.

\begin{figure}[!t]
\begin{center}\includegraphics[width=4.5in]{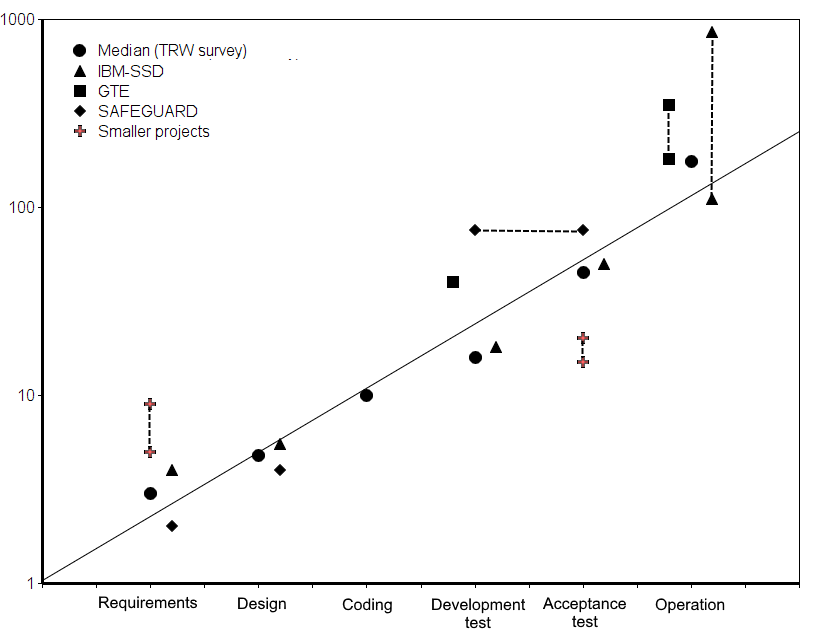}\end{center}
 \caption{Historical cost-to-fix curve. Adapted from~\cite{Boehm81}, p. 40.}\label{fig:cost-to-fix}
 \end{figure}

Figure~\ref{fig:cost-to-fix} shows the DIE as reported in \textit{Software Engineering Economics}~\cite{Boehm81} based on data from large systems in the late 70s from IBM~\cite{Fagan76}, TRW~\cite{Boehm76}, GTE~\cite{Daly77}, and Bell Labs~\cite{Stephenson76}. 
We note that it is unclear from the text in~\cite{Daly77} and \cite{Boehm76} if cost is defined in terms of effort, or in actual cost (i.e., labor, materiel, travel, etc). The data points from these studies are not published for analysis.
Baziuk~\cite{baziuk1995bnr} reports an exponential increase in the cost to patch software in the field versus system test, and Stecklein et al.~\cite{steck04} produce a cost-to-fix curve (as price) that fits precisely with \fig{cost-to-fix}. Westland~\cite{westland2002cost} finds that the cost to fix engineering errors is exponentially related to the cost of the overall cost of a case study project. Reifer~\cite{reifer2007profiles} confirms the exponential increase in the DIE in 19 CMMI Level 5 organizations though this appears to be based on survey rather than empirical data. 


Shull et al.~\cite{Shull02} conducted a literature survey and held a series of e-workshops with industry experts on fighting defects. Workshop participants from Toshiba and IBM reported cost-to-fix ratios between early lifecycle and post-delivery defects of 1:137 and 1:117 for large projects respectively~\cite{Shull02} -- but the raw data points were not provided and thus cannot be confirmed. 
Elssamadisy and Schalliol~\cite{Elssamadisy02} offer an anecdotal report on the growing, high cost of rework in a 50 person, three-year, 500KLOC Extreme Programming project as the project grew in size and complexity-- but again we cannot access their 
exact figures. This was a common theme in the literature reviewed for this paper-- i.e.  that  it was no longer possible to access the data used to make prior conclusions.



 

Some studies report smaller
 increases in the effort required to fix delayed issues.  Boehm~\cite{Boehm80} provides data suggesting that the cost-to-fix curve for small projects
 is flatter than for large projects (the dashed line of Figure~\ref{fig:cost-to-fix}). Data from NASA's Johnson Space Flight Center, reported by Shull~\cite{Shull02}, found that the cost to fix certain non-critical classes of defects was fairly constant across lifecycle phases (1.2 hours on average early in the project, versus 1.5 hours late in the project). Royce~\cite{Royce98} studied  a million-line, safety-critical missile defense system. Design changes (including architecture changes) required approximately twice the effort of implementation and test changes, and the cost-to-fix in implementation and test phases increased slowly. Boehm~\cite{Boehm10} attributes this success to a development process focused on removing architecture risk early in the lifecycle. Willis et al. (\cite{willis1998hughes}, page 54) provide tables summarizing the effort to fix over 66,000 defects as a function of lifecycle phase injected and removed from multiple projects. The tables are partly obscured, but seem to provide the first large scale evidence that a)~DIE need not be exponential and b)~DIE need not be monotonically increasing.  Again, the data points from these studies are not available, and thus newer evidence both in favor of and contrary to the DIE cannot be evaluated.

To gain a sense of how current the perception of the DIE is, 
we conducted two surveys of software engineers. 
The surveys collected data on software engineers' views of the DIE and other commonly held software engineering ``laws''. The surveys were conducted using Amazon's Mechanical Turk. The first survey was conducted only with professional software engineers. Participants were required to complete a pretest to verify their status as a professional or open source software developer and to confirm their knowledge of basic software engineering terminology and technology. The  second survey was conducted with Program Committee members of the ESEC/FSE 2015 and ICSE 2014 conferences solicited via email.

The practitioner survey presented the following law: ``requirements errors are the most expensive to fix when found during production but the cheapest to fix early in development'' (from Glass~\cite{glass02} p.71 who references Boehm \& Basili~\cite{boehm01}). We abbreviate this law as RqtsErr.\footnote{We use the RqtsErr formulation since this issue typically needs no supportive explanatory
text. If we had asked respondents about our more general term ``delayed issue
effect'', we would have had to burden our respondents with extra explanations.}
The PC member survey presented the RqtsErr law and an additional law on the DelayedIssueEffect: ``In general, the longer errors are in the system (requirements errors, design errors, coding errors, etc.), the more expensive they are to fix''.  The respondents answered two questions in response to each law:
\bi
\item \textbf{Agreement:} ``Based on your experience, do you agree that the statement above is correct?'' A Likert scale captured the agreement score from Strongly Disagree to Strongly Agree. A text box was provided to explain the answer. 
\item \textbf{Applicability:} ``To the extent that you believe it, how widely do you think it applies among software development contexts?'' The possible answers were: -1: I don't know, 0: this law does not apply at all, ..., 5: always applies. Respondents were required to explain the applicability score in a text box.
\ei

\begin{figure}[ht] 
\scriptsize 
 \begin{center}
\begin{tabular}{l|c|c|c|c|c}
 &  & \multicolumn{2}{c|}{agreement} & \multicolumn{2}{c}{applicability} \\
Practitioner survey  & N & med & mode & med & mode \\
\hline 
\textbf{Rqts errors are most expensive...} & 16 & 5 & 5 & 4 & 5 \\ 
Inspections can remove 90\% of defects & 18 & 4 & 5 & 4 & 5 \\
80-20 rule (defects to modules) & 12 & 4 & 5 & 4 & 5 \\
Most time is spent removing errors & 16 & 4 & 4 & 4 & 5 \\ 
Process maturity improves output & 17 & 4 & 4 & 4 & 4 \\ 
Missing reqts are hardest to fix & 17 & 4 & 4 & 4 & 4 \\
Reuse increases prod. and qual. & 16 & 4 & 4 & 4 & 4 \\
OO-programming reduces errors & 13 & 4 & 4 & 4 & 3 \\
Adding manpower to a late project & 15 & 4 & 4 & 4 & 4 \\
Smaller changes have higher error density & 14 & 3 & 3 & 3.5 & 5 \\
A developer is unsuited to test own code & 17 & 3 & 1 & 4 & 5\\
 \multicolumn{6}{l}{}\\
\multicolumn{6}{l}{Researcher survey} \\\hline 
Process maturity improves output & 4 & 4 & 4 & 4 & 5 \\
\textbf{Rqts errors are most expensive...} & 30 & 4 & 4 & 4 & 4   \\ 
\textbf{DelayedIssueEffect} & 30 & 4 & 4 & -- & --  \\
Reuse increases prod. and qual. & 6 & 4 & 4 & 4 & 4 \\
80-20 rule (defects to modules) & 6 & 4 & 4 & 4 & 3 \\
Missing reqts are hardest to fix & 7 & 4 & 4 & 4 & 3 \\
OO-programming reduces errors & 6 & 4 & 4 & 3 & 4 \\
Inspections can remove 90\% of defects & 7 & 4 & 4 & 3 & 3 \\
Adding manpower to a late project & 4 & 3 & 4 & 4 & 3 \\
Most time is spent removing errors & 6 & 3 & 3 & 4 & 4 \\ 
Smaller changes have higher error density & 4 & 3 & -- & 4 & 4 \\
A developer is unsuited to test own code & 7 & 2 & 1 & 3 & 3
\end{tabular} 
 \end{center}
\caption{Agreement and applicability of SE axioms.}
\label{fig:survey_results}
\end{figure}

Summary statistics for the agreement and applicability scores for the RqtsErr and DelayedIssueEffect laws are presented in Figure~\ref{fig:survey_results}. Responses whose Applicability response was ''I don't know'' are omitted from analysis. Laws other than RqtsErr and DIE are not relevant to this paper, but are shown for comparison.

Both  practitioners and researchers strongly believed in RqtsErr. In both sets of responses, RqtsErr received  scores higher than most
other laws. Overall, the RqtsErr law was the most agreed upon and most applicable law of 11 surveyed amongst practitioners, and the second most agreed upon law amongst researchers. From the free response texts, we note that the researchers who disagreed with RqtsErr generally asserted that requirements change can be expensive, but that the effect depends on the process used (e.g., agile vs. waterfall) and the adaptability of the system architecture.

The above arguments provide evidence to the claim that the DIE is both poorly documented yet (still) widely believed. The comments of Glass~\cite{glass02}, that the DIE is ``just common sense'', suggest that DIE may be the target of confirmation bias. An example of this is \fig{steck} from~\cite{steck04}, which purports to show nine references to ``studies [that] have been performed to determine the software error cost factors''. Only one of these sources, \textit{Software Engineering Economics}~\cite{Boehm81}, is based on real project data.  Despite a lack of recent evidence, the perception of the DIE persists today among both the software engineers sampled in our survey and in popular literature. In the intervening years, many advances in software technology and processes have been made precisely to deal with risks such as the DIE. Thus, it is appropriate to ask the question, does the DIE still exist?

 \begin{figure}[!ht] 
{\small
\begin{center}
\begin{tabular}{r|rrrr|l}
                        & \multicolumn{4}{c}{Phase Requirements Issue Found }   &     \\
 Cited source           & Requirements  &   Design  &   Code    &  Test         & Data sources used to determine DIE          \\\hline
\cite{Boehm81}          &   1           &   5       &   10      &   50          & Multiple projects     \\  
Hoffman, 2001           &   1           &   3       &   5       &   37          & Unknown - no bibliography entry \\ 
\cite{Cigital2003}      &   1           &   3       &   7       &   51          & Extrapolated from defect counts for a one project \\
\cite{Rothman2000}      &               &   5       &   33      &   75          & Fictitious example \\
\cite{Rothman2000} Case B   &               &           &   10      &   40      & Fictitious example \\
\cite{Rothman2000} Case C   &               &           &   10      &   40      & Fictitious example \\
\cite{Rothman2002}      &   1           &   20      &   45      &   250         & Fictitious example \\
\cite{Pavlina2000}      &   1           &   10      &   100     &   1000        & None provided \\
\cite{McGibbon2007}     &               &   5       &           &   50          & Pen \& paper exercise - no real data \\
\end{tabular}
\end{center}}
\caption{Confirmation bias -- sources for DIE cited in Table 1 of~\cite{steck04}. Note that \textit{all} of these are cited as ``studies [that] have been performed to determine the software error cost factors'', but only one, \cite{Boehm81}, is backed by actual data.}\label{fig:steck}
\end{figure}


\subsection{Early Onset of the DIE Effect}\label{sect:earlyonset}

One feature of the the DIE literature is important to our subsequent discussion:   
the onset of  DIE  {\em prior} to delivery.
\bi
\item  
\fig{b81} reports a 40-fold increase in effort requirements to acceptance testing
\item
 \fig{cost-to-fix} reports a 100-fold increase
(for the larger projects) before the code is delivered 
\ei
Any manager noticing this  early onset of DIE (prior to delivery, during the initial development)
would be well-justified
in believing that  the difficulty in resolving issues  will get much worse. Such managers
would therefore expect DIE to have a marked effect post-deployment.
We make this point since,  in the new project data presented below, we focus on DIE pre-delivery.

\section{Delayed Issues are not  Harder  to Resolve}
\label{sect:analysis}
The above analysis motivates a more detailed look at the delayed issued effect.  
Accordingly, we examined 171 software projects conducted between 2006 and 2014. 

These projects took place at organizations in many countries and were conducted using  the Team Software Process (TSP$\textsuperscript{SM}$). Since 2000, the SEI has been teaching and coaching TSP teams. One of the authors (Nichols) has mentored software development teams and coaches around the world as they deploy TSP within their organizations since 2006.  The  most recent completions were in 2015.

\begin{figure}
\begin{center} 
\includegraphics[width=2.5in]{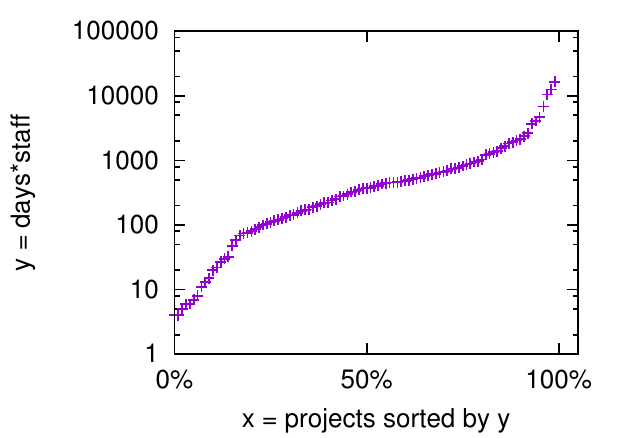}
\end{center} 
\caption{Distribution of 
{\em effort} (which is
{\em team size} times
{\em days of work}). For example, if 10 programmers work for 10 days,
then the effort is 100 days. The median value in this plot
 271 days.}\label{fig:dist}
\end{figure}

The projects were mostly small to medium, with a median duration of 46 days and a maximum duration of 90 days in major increments. 
Several projects extended for multiple incremental development cycles. 
Median team size was 7 people, with a maximum of 40. See \fig{dist} for the total
effort seen in those projects. Many of the projects were e-commerce web portals or banking systems in the US, South Africa, and Mexico. 
There were  some  medical device projects in  the US, France, Japan, and Germany as well  as a commercial computer-aided design systems, and embedded systems. A more thorough characterization of the projects providing data is provided in \S\tion{data_character}.

An anonymized version of that data is available in the PROMISE repository at openscience.us/repo.
For confidentiality restrictions, we cannot offer 
further details on these projects.

\subsection{About TSP$\textsuperscript{SM}$}\label{sect:tsp}

TSP is a software project management approach developed at the Software Engineering Institute (SEI) at Carnegie Mellon University~\cite{tsp00}. TSP is an extension of the Personal Software Process (PSP$\textsuperscript{SM}$) developed at the SEI by Watts Humphrey~\cite{tsp00}.


Common features of TSP projects include {\em planning}, {\em personal reviews}, {\em peer inspections}, and {\em coaching}.
A TSP {\em coach} helps the team to plan and analyze performance. The coach is the only role authorized to submit project data to the SEI.
Before reviewing data with the teams, therefore before submission, these coaches check the data for obvious errors.

During {\em Planning}, developers estimate the size of work products and convert this to a total effort using historical rates. Time in specific tasks come from the  process phases and historical percent time in phase distributions. Defects are estimated using historical phase injection rates and phase removal yields. Coaches help the developers to compare estimates against actual results. In this way, developers acquire a more realistic understanding of their work behavior, performance, and schedule status.

{\em Personal review} is a technique taken from the PSP and its use in TSP is unique.  Developers follow a systematic process to remove defects by  examining their own work products using a checklist built from their personal defect profile. This personal review occurs after some product or part of a product is considered to be constructed and before peer reviews or test. 


{\em Peer inspection} is a  technique in
traditional software engineering and is often called peer review.
 Basili and Boehm   commented in 2001~\cite{boehm01} 
that peer reviews can catch over half the defects introduced into a system.
Peer inspection can be conducted on any artifact generated anywhere in the software
lifecycle and can quickly be adapted to new kinds of artifacts. TSP peer reviews follow the Fagan style in which the reviewer uses a checklist composed of common team defects prior to a review team meeting. 
 
Overall, the   effort associated with adding TSP to a project is not onerous. McHale reports~\cite{mchale02}:
\bi
\item
 The time spent  tracking time, defects, and tasks requires less than 3\% of a developer's time. Weekly team meetings  require at most an hour, which is
only 2.5\% of a 40 hour work week. 
\item
Team launches and replans average about 1 day per month or 5\% planning overhead.
\ei
It is true that one staff member is needed as a ``coach'' to mentor the teams
and certify and monitor that data collection. However, one of us (Nichols) has worked with dozens of TSP teams. He reports that one  trained coach can support 4 or 6 teams (depending upon team experience).

\subsection{Data Collection and Definitions}
\label{sect:data-collection}

Organizations using TSP agree to provide their project data to the SEI for use in research. In return the SEI agrees that  data must not be traceable to its source. The data are collected at major project events: launch, interim checkpoints, and at project completion. The data from these TSP projects were collected and stored in the Software Engineering Measured Process Repository (SEMPR) at the SEI. 

As of November 2014, the SEI TSP database contained data from 212
TSP projects. The projects completed between July 2006 and
November 2014; they included 47 organizations and 843 people. 
The database fact tables
contain 268,726 time logs, 
154,238 task logs,
 47,376 defect logs, 
and 26,534 size logs. 
In this paper, we exclude 41 of the 212 that had too few defects (less than 30), leaving 171 projects included in the analysis.

\subsubsection{Definition: time for plan item}

Using a tool supporting the SEI data specification, developers keep detailed time-tracking logs. The time-tracking logs record  work start time, work end time,  delta
work time, and interruption time. Software engineers are often
interrupted by meetings, requests for technical help, reporting, and
so forth. These events are recorded, in minutes, as interruption
time. In TSP, time logs are recorded against \textit{plan items}. A planned item is a specific task assigned to a specific developer, such as resolving a defect, coding a feature, performing an inspection or writing a test. Each  work session includes a start time, an end time, and interruption time. The active time, or actual time for the plan item is calculated by summing the active time durations for all work sessions on that task.
\[
\text{\emph{actual time for plan item}} := \text{SUM(}\text{end time} - \text{start time} - \text{interruption time}) 
\]
Time is tracked per person per plan item in the time-tracking logs, e.g. a 30 minute design review session involving 3 people will have three time log entries summing to 90 minutes. Time includes the time to analyze, repair, and validate a defect fix.


\subsubsection{Definition: defects and time-to-fix}\label{sect:defin}
In the TSP, a  \emph{defect} is any change to a product, after its construction, that is necessary to make the product correct.  A typographical error found in review is a defect. If that same defect is discovered while writing the code but before review, it is not considered to be a defect. 
SEI TSP defect types are:
\bi
\item Environment: design, compile, test,  other support  problems
\item Interface: procedure calls and reference, I/O, user format
\item Data: structure, content
\item Documentation: comments, messages
\item Syntax: spelling, punctuation typos, instruction formats
\item Function: logic, pointers, loops, recursion, computation  
\item Checking: error messages, inadequate checks
\item Build: change management, library, version control
\item Assignment: package
declaration, duplicate names, scope
\item System: configuration, timing, memory
\ei

\begin{figure}[t]
\begin{center}
\includegraphics[width=2.5in]{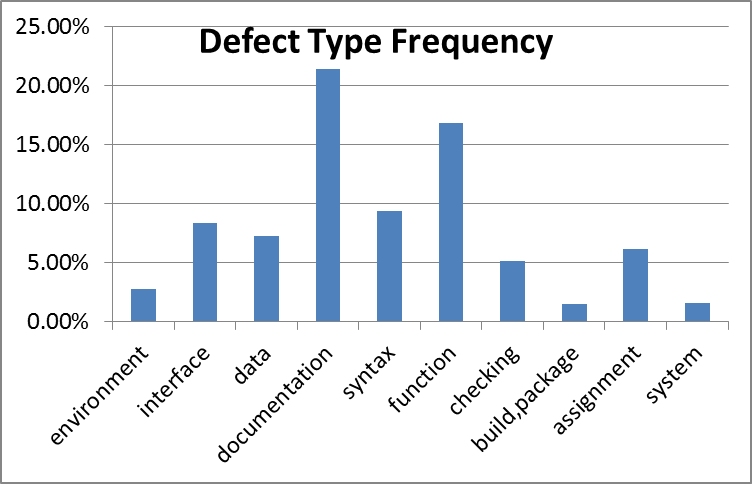}
\end{center}
\caption{Relative frequencies of these defect types seen in our TSP data.}\label{fig:dtypes}.
\end{figure}
In our TSP data,  the relative frequencies of these defect types are shown in \fig{dtypes}. Around a quarter of the fixes were simple documentation changes.
That said, 75\% of the changes are quite elaborate;   e.g. fixes to function necessitates a careful reflection of the purpose of the code.

Individual defects are recorded as line items in the defect logs uploaded to the SEMPR at the SEI. The defect entry includes the time and date a defect was discovered, the phase in which that defect was injected, the development phase in which it was removed, the time (in minutes) required to find and fix the defect, and the categorical type. 

In the TSP, defect data includes the affected artifact, the estimated developer fix effort (find and fix), the lifecycle phases in which the defect was injected and removed, and the developer who implemented the fix. In the database, the task is associated with a plan item. Defects (one or more) are recorded in the defect log and associated with the plan item (task) in the time tracking logs. For example, a review session, an inspection meeting, or a test would be plan items associated with some product component. When defects are found and fixed, the time recorded in the time-tracking logs against the plan items includes the direct effort time (stop watch rather than wall clock time) required to (a)~collect data and realize there is an error, 
(b)~prepare a fix,  and (c)~apply some validation
procedure to check the fix (e.g. discuss it with a colleague or execute some tests). Although we have explicit estimates of "find and fix" effort for each defect, this fails to account for the full costs (e.g. meeting time or test execution). Because the vast majority of defects are removed in explicit removal phases, we chose to estimate defect cost using the entire time in removal phases divided by the number of defects. We recognize that this approach can exaggerate cost per defect for cases with few defects and large overhead effort, such large test suites or slow running tests that require continuous developer attention.  Nonetheless, this approach provides a better comparison between early removals from inspections later removals from test. The result will be a time per defect that is greater than the directly measured "find and fix" time, but smaller than the wall clock or calendar time. 

Since multiple defects can be recorded against a plan item, the time-to-fix a defect is defined as:
\[
\text{\emph{time-to-fix a defect}} := \frac{\text{time for defect plan item}}{\text{\# of defects in plan item}}
\]

\subsubsection{Definition: development phase} \label{development_phase}
\begin{figure}[!b]  
\begin{center}
\includegraphics[width=1.6in]{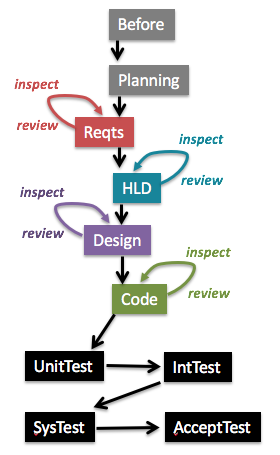}  
\end{center}
\caption{Phases of our data.
Abbreviations: 
{\em Before}= before development; 
{\em Reqts}	  = requirements; 
{\em HLD}	  = high-level design; 
{\em IntTest} = Integration testing (with code from others); 
{\em SysTest} = system test (e.g. load stress tests); 
{\em AcceptTest}  = acceptance testing (with users); 
{\em review}        = private activity; 
{\em inspect}        = group activity.}
\label{fig:waterfall}
\end{figure}

The \textit{development phases} against which plan items are logged in the data are shown in \fig{waterfall}.
Although the representation suggests a waterfall model, the SEI experience is that the projects follow a spiral approach or perform the work in iterative and/or incremental development cycles. The phases are thus the logical stages through which each increment must progress during development.

One special feature of  \fig{waterfall} is the {\em before} phase, in which the TSP team assures that management has clearly identified cost, schedule, and scope goals appropriate to the upcoming development activities, often including a conceptual model~\cite{Humphrey:2005}. For example an architecture team must have sufficient requirements to reason about, prototype, and specify an architecture~\cite{Bachmann13} while a coding only team within a larger project would have more precisely defined requirements and high level design.
 

Note that, in Figure~\ref{fig:waterfall}, several  phases in which the product is created have sub-phases of {\em review} and {\em inspect} to remove defects. As discussed in \S\ref{sect:tsp}, individuals perform personal reviews of their work products prior to the peer review (which TSP calls the inspection). Testing activities are divided as follows. Developers perform unit test prior to code complete.  After code complete a standard phase is integration, which combines program units into a workable system ready for system test. Integration,  system test, and acceptance test are often performed by another group.

\subsection{Data Integrity}
\label{sect:data_integrity}

A common property of real-world data sets is the presence
of noisy entries (superfluous  or spurious data). 
The level of noise can be quite high. For example, as reported
in \cite{shepperd12}, around
10\% to 30\%
of the records in the NASA MDP defect data sets are
affected by noise. 

One reason to use the SEI data for the analysis of this paper is its remarkably low level of noise.
Nichols et al.~\cite{shirai14}  report that
the noise levels in the SEI TSP data are smaller than those seen
in other data sets. They found in the SEI TSP data that:\bi 
\item
4\% of the data was incorrect (e.g. nulls, illegal formats);
\item  2\% of the data has inconsistencies such as timestamps
where the stop time was before the start time;
\item 3\% of the data contained values that were not credible
such as tasks listed in one day that took more than six hours for a single developer.
\ei 
One explanation for this low level of noise is the TSP process.
One the guiding principles of TSP was that  people performing the work are  responsible for planning and tracking the work. That is,  all the data collected here was entered
by local developers, who use the data for planning and tracking their projects. This data was then checked by local coaches before being sent to the SEI
databases. While coaches are certified by demonstrating competent use of the TSP process with the artifacts and data,  project success or performance is not a criterion. 
The use of certified local coaches within each project increases the integrity of our data.

\subsection{Project Descriptive Characteristics}
\label{sect:data_character}

In this section we provide some descriptive statistics, discuss the projects from which this data was drawn, summarize some additional contextual information. The project contexts describe the conditions under which these measures were obtained, help determine relevance of the results, and may guide future data analysis with segmentation. Key attributes of the context include the business and application domains, product size, project duration,  work flows, team size, team management,  development and integration approaches, organization size, location or distribution, certifications, developer experience, programming languages and tools used. 

We are unable at this time to provide all individual context data for each of the projects for several reasons. While the development data was recorded in tools and submitted in a structured form, context data was collected in less structured project questionnaires, site questionnaires, team member surveys, launch presentations and reports, post mortem presentations and reports. This data has not yet been mined from the submissions 1) because of the cost and effort required, 2) we are obligated to avoid providing any data that can identify projects (that is, the data must remain anonymous), and 3) the unstructured data may not be complete when submitted.   Gathering more projects will make it easier to anonymize the data and overcome missing data problems. Interest in the data sets by the community may encourage our sponsor to fund additional data mining.  Nonetheless, much context is available from the project data  and we provide some additional context not included within the fact sheets. 
 
\begin{figure}[!t] 
\begin{center}
\includegraphics[height=1.5in]{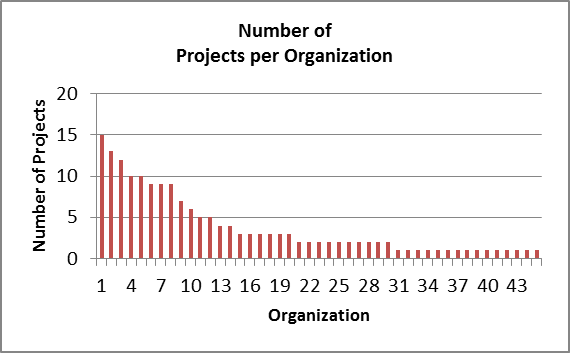}
\end{center} 
\caption{Number of projects per development organization.}
\label{fig:number of projects per organization}
\end{figure}

 The projects included come from 45 unique organizations from 6 countries. \fig{nationality} shows the country of origin and application domains for the projects.  \fig{number of projects per organization} shows the number of projects from each organization. 
 
The most common countries of origin are the US and Mexico. Not apparent in this display is that the US companies tend to be fewer and larger with many projects while the Mexican companies are more likely to have one to several projects. Several companies, typically larger companies, are international with development teams in the US and either France or China. 

The most common project application domains are banking, consumer applications, engineering design tools, and medical devices.   
The data for programming languages is incomplete, with most projects using more than one language, but few reporting  programming language by component or size.  The list of languages includes ABAP, ADA, Alpha, C, C++, C\#, ASP.net, Delphi, Gauss, Genexus, Hotware, HTML, Java, JavaScript, PHP, PLSQL, Ruby, SQL, and Visual Basic.

 The specific process work flows and practices are developed by the development team personnel who have received specific training on defining work processes as part of their Personal Software Process training. The process data was collected by the team members to self-manage their personal and team work. The members also exhibited self-management behavior by estimating planning and scheduling the work tasks. 

While the processes and work flows among these projects can vary, the logical order described in section \ref{development_phase} is followed. Development tasks such as requirements development, design, or code, are typically followed by an appraisal phase such as personal review or inspection. Effort and effectiveness of these activities vary among projects and developers.

\begin{figure}[ht]
\scriptsize
\centering
\begin{tabular}{cc}
\begin{tabular}[t]{lr}
Country      & \% of projects \\\hline
China        & 1.0 \%            \\
France       & 10.0 \%           \\
Mexico       & 41.0 \%           \\
South Africa & 4.0 \%            \\
UK           & 1.5 \%          \\
US           & 42.5 \%        
\end{tabular}
&
\begin{tabular}[t]{lr}
Application Domain         & count \\\hline
Aviation                   & 19    \\
Banking                    & 23    \\
Business intelligence      & 19    \\
Construction Support tools & 3     \\
Consumer applications      & 24    \\
Custom Applications        & 1     \\
Embedded systems           & 2     \\
Engineering Design Tools   & 21    \\
Geography and Mapping      & 2     \\
Government                 & 5     \\
Human Resources Management & 3     \\
Information Technology     & 1     \\
Manufacturing              & 3     \\
Medical Devices            & 15    \\
Other                      & 2     \\
Payroll services           & 1     \\
Solutions Integration      & 1     \\
Web applications           & 13    \\
Wholesale or retail trade  & 9    
\end{tabular}
\end{tabular}
\caption{Project nationality and application domain}
\label{fig:nationality}
\end{figure}

 \begin{figure}[ht]
\scriptsize
\centering
\begin{tabular}{llr} 
 Begin Phase &  Final Phase & Count  \\\hline
  Requirements      & Unit Test                   & 12  \\ 
  Requirements      & Build and Integration Test  & 19  \\ 
  Requirements      & System Test                 & 36  \\ 
  High Level Design & Unit Test                   & 3  \\ 
  High Level Design & Build and Integration Test  & 10  \\ 
  High Level Design & System Test                 & 9  \\ 
  Detailed Level Design & Unit Test                   & 18  \\ 
  Detailed Level Design & Build and Integration Test  & 24  \\ 
  Detailed Level Design & System Test                 & 17  \\ 
\end{tabular}
\caption{Earliest and latest process phases used by the projects}
\label{fig:earliest-and-least-process-phases}
\end{figure}

\begin{figure}[ht]
\scriptsize
\centering
\renewcommand{\arraystretch}{1}%
\begin{tabular}{lrrrrrrrl}
                                        & N   & Min & Q1   & Median & Q3    & Max   & Mean   &  Distribution \\\hline
Team size                               & 171 & 1   & 4    & 6      & 10    & 36    &  7.8    &  \includegraphics[width=25mm]{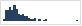} \\ 
Duration {[}days{]}                     & 171 & 7   & 33   & 61     & 118   & 1918  & 107    & \includegraphics[width=25mm]{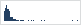} \\
Added \& Modified LOC                   & 117 & 2   & 1125 & 4201   & 13092 & 88394 & 10259  & \includegraphics[width=25mm]{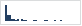} \\
Defects Found \& Fixed                  & 171 & 1   & 28   & 95     & 278   & 4580  & 324.4  & \includegraphics[width=25mm]{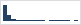} \\
Components                              & 171 & 8   & 26   & 49     & 107   & 4170  & 116.2  & \includegraphics[width=25mm]{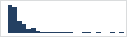} \\
Project Initiation Year                 & 171 &    2006 & 2011 & 2012   & 2013  & 2014  & 2011.9 & \includegraphics[width=25mm]{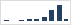}
\end{tabular}

\caption{Project summary description}
\label{fig:Project-descriptive-stats}
\end{figure}

 The project schedule, cost, and scope  are characterized by calendar duration, development team size project, and product size (measured in added and modified lines of code and number of components).  These data are all available from the project fact sheets for each project. Summary statistics and the year of project initiation are displayed in Figure~\ref{fig:Project-descriptive-stats}. From this table we can make some observations about the range of project characteristics.
 
 Of the 171 projects in the sample, only 117 collected size data in lines of code. However all projects tracked effort and  the component counts with applied effort are provided. Other data are complete for all 171 projects. The projects were mostly of short duration and small to medium size. The median project began in 2012 lasted 61 days, produced 4,200 Lines of Code, 49 components (modules or features). Duration ranged from 7 to 1,918 days. Size ranged from minimal (this may represent a short maintenance project) to 88,394. The earliest project was in 2006 and the most recent in 2014. 
 
How many of these teams could be classified as "agile" is not clear because actual practices in the agile world can vary. We did not ask teams to self-identify, however we offer the following observations regarding characteristics commonly associated with agile behavior. 
\begin{itemize}
    \item all teams were self managed, defining work flows, practices, and schedules
    \item teams met at least weekly to evaluate progress and re-plan
    \item most teams were small with a median size of 6 and a mean of 7.8; only 25\% of the teams were larger than 10 with a long tail on the distribution
    \item the median project lasted only 60 days, suggesting limited scope for each integration
\end{itemize}

\begin{figure}[!t] 
\vspace{0.5cm}
\begin{center}
\includegraphics[height=2.5in]{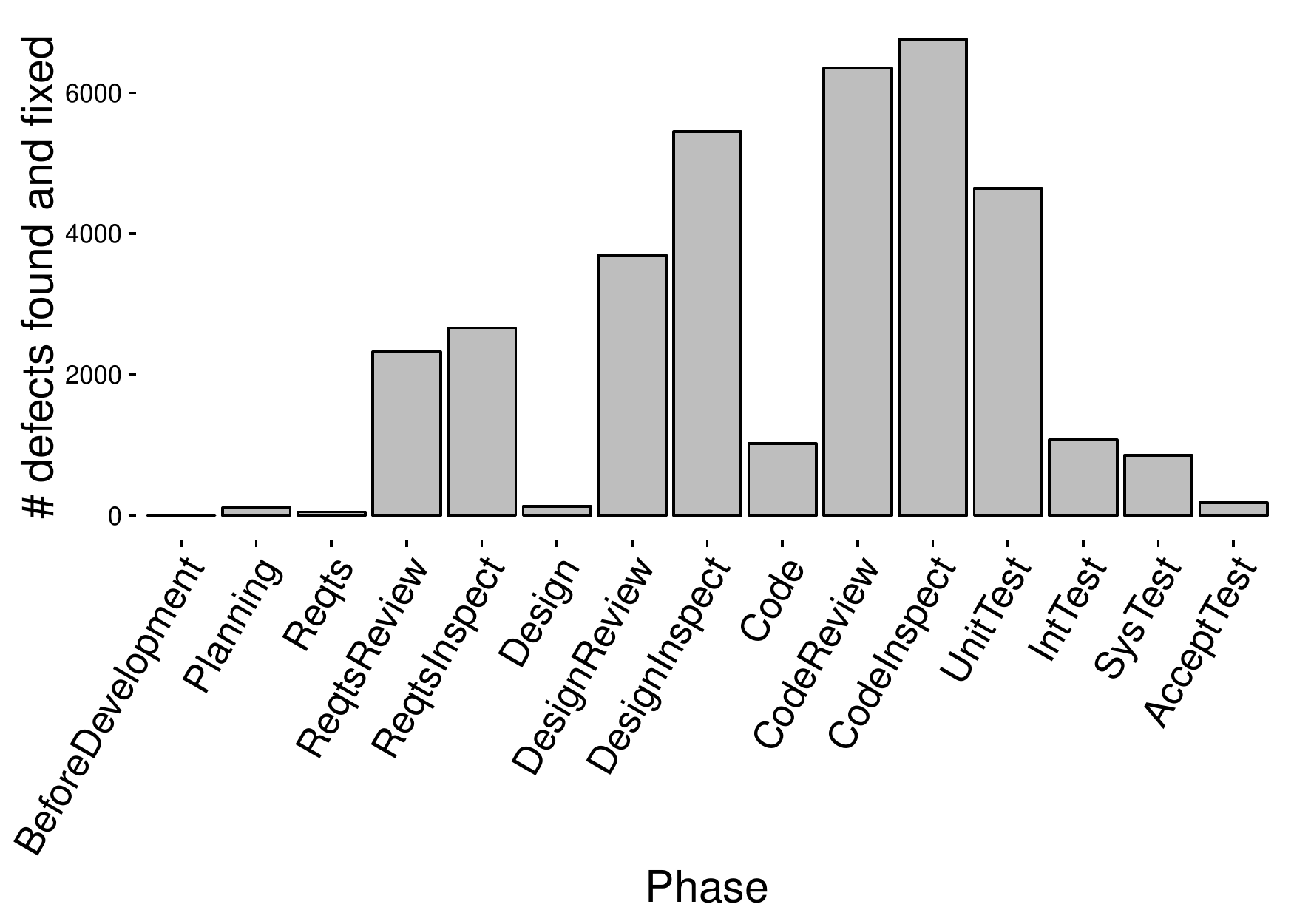}
\end{center} 
\caption{Distribution of defects by phase removed.}
\label{fig:fix-phase-dist}
\end{figure}

\subsection{Statistical Analysis}\label{sect:stats}
    
    In the following presentation of our results, three statistical methods were used to test for the delayed
    issue effect: the Scott-Knott ranker;   bootstrap sampling (to test for statistical significantly
    different results); and an effect size test (to reject any significant differences that are trivially small).
    Scott-Knott allows for a direct answer to the following questions:
    \bi
    \item
    Given an issue
    raised at phase $i$ and resolved at phase $\forall j,k \in \{i,i+1,i+2,...\}$,...
    \item 
    ... Is it
    true that  the time to resolve issues in phase $j$ is significantly different 
    to the time to resolve issues in phase $k$? 
    \ei
    Note that if $j,k$ times are significantly different, then we can compare
    the median values to say (e.g.) resolution time at phase $k$ is 3 times slower than phase $j$.
    Note also that if all times $j,k$ are {\em not} significantly different
    then we say that the phases all {\em rank} the same (and we denote this by setting
    all such {\em ranks} to 1).
    
    In the following results, we nearly always encountered the second case;
    i.e.   the times to resolve issues at different times were usually {\em not significantly}
    different.
    
    As to technical details of the Scott-Knott methods, this ranker 
      was   recommended  by  Mittas and Angelis in a
    recent TSE'13 article~\cite{mittas13} and by Ghotra et al. in a recent
    ICSE'15 article~\cite{ghotra2015icse}
    Scott-Knott is a 
    top-down clustering approach used to rank different treatments. If that 
    clustering finds an ``interesting division'' of the data, then some 
    statistical test is applied to the two divisions to check if they are 
    statistically significant different. If so, Scott-Knott considers recurses into both 
    halves.
    Before Scott-Knot recurses,  however,  it applies some  statistical hypothesis test $H$ to check
    if $m,n$ are significantly different. 
    To operationalize ``interesting'', 
    \bi
    \item
    Scott-Knott seeks the division of 
    $l$ treatments into subsets of size $m,n$ (so if $n$ was 
    appended to the  end of $m$ then that new list
    would the same as $l$).
    \item
    We say that $l,m,n$ have
    sizes $ls,ms,ns$ and median values
    $l.\mu, m.\mu, n.\mu$ (respectively) 
    \item
    Scott-Knott tries all ways to split $l$ into $m,n$ and returns
    the one that maximizes the differences in the mean values
    before and after the splits; i.e. 
 \[\frac{ms}{ls}abs(m.\mu - l.\mu)^2 + \frac{ns}{ls}abs(n.\mu - l.\mu)^2\]
   \ei
   To operationalize $H$, we use both  bootstrap sampling  and Vargha and Delaney's A12 effect size test.
   In other 
    words, we divide the data if \textit{both} bootstrap sampling and effect 
    size test agree that a division is statistically significant (with a 
    confidence of 99\%) and not a small effect ($A12 \ge 0.6$).
    For a justification of the use of non-parametric bootstrapping, see Efron 
    \& Tibshirani~\cite[p220-223]{efron93}. For a justification of the use of 
    effect size tests see Shepperd and MacDonell~\cite{shepperd12a}; 
    Kampenes~\cite{kampenes07}; and Kocaguenli et 
    al.~\cite{Kocaguneli2013:ep}. These researchers warn that even if a 
    hypothesis test declares two populations to be ``significantly'' 
    different, then that result is misleading if the ``effect size'' is very 
    small. Hence, to assess the performance differences we first must rule out 
    small effects using Vargha and Delaney's A12 test, a test    endorsed by Arcuri and 
    Briand at ICSE'11~\cite{arcuri11}.

    To  apply Scott-Knott, we divided data into the phases $P_0$ where issues are introduced. Next, for each division, we separated all the issues that were removed at different subsequent issues $P_r \in \{P_1,P_2,..\}$.
    For each pair $P_0,P_r$, we build one treatment containing the  issue resolution times for     issues raised in $P_0$ and resolved in $P_r$. These treatments
    were then ranked by Scott-Knott.

\begin{figure} 
\begin{center}

\scriptsize
\begin{tabular}{c|lr|rr|rl}
& \multicolumn{2}{c|}{ } & \multicolumn{2}{c|}{Percentiles} & \multicolumn{2}{c}{~ } \\ 
  & \multicolumn{2}{c|}{ } & \multicolumn{2}{c|}{(units = } & \multicolumn{2}{c}{Growth with respect to earliest phase  } \\ 
   & \multicolumn{2}{c|}{Phase} & \multicolumn{2}{c|}{  minutes)} & \multicolumn{2}{c}{(unitless  ratios of two time values) } \\\hline

  rank & injected & removed & 50th &   IQR & \multicolumn{2}{c}{50th percentile growth}  \\ 
\hline

\multicolumn{7}{c}{~}  \\
1 &  Before   
      &   DesignInspect   & 10 &  14 & 1.00 & \textcolor{black}{\rule{10mm}{2mm}}  \\
1 &   &   CodeReview      & 8 &  14 & 0.80 & \textcolor{black}{\rule{8mm}{2mm}}  \\
1 &   &   CodeInspect     & 10 &  16 & 1.00 & \textcolor{black}{\rule{10mm}{2mm}}  \\
1 &   &   UnitTest        & 12 &   21 & 1.20 & \textcolor{black}{\rule{12mm}{2mm}}  \\
1 &   &   IntTest         & 15 &   31 & 1.50 & \textcolor{black}{\rule{15mm}{2mm}}  \\
1 &   &   SysTest         & 11 &   22 & 1.10 & \textcolor{black}{\rule{12mm}{2mm}}  \\  
\hline\multicolumn{7}{c}{~}  \\
1 &  Planning     &   ReqtsReview     & 8 &  14 & 1.00 & \textcolor{black}{\rule{10mm}{2mm}}  \\
1 &               &   DesignInspect   & 11 &  13 & 1.38 & \textcolor{black}{\rule{13mm}{2mm}} \\
2 &               &   UnitTest        & 24 &   25 & 3.00 & \textcolor{black}{\rule{30mm}{2mm}} \\
\hline\multicolumn{7}{c}{~}  \\
1 &  Reqts   &   ReqtsReview     & 13 &  20 & 1.00 & \textcolor{black}{\rule{10mm}{2mm}} \\
1 &          &   ReqtsInspect    & 12 &   18 & 0.92 & \textcolor{black}{\rule{9mm}{2mm}}  \\
1 &          &   DesignReview    & 10 &   14 & 0.77 & \textcolor{black}{\rule{7mm}{2mm}} \\
1 &          &   DesignInspect   & 9 &   15 & 0.69 & \textcolor{black}{\rule{6mm}{2mm}}   \\
1 &          &   CodeInspect     & 13 &  24 & 1.00 & \textcolor{black}{\rule{10mm}{2mm}} \\
1 &          &   UnitTest        & 10 &  17 & 0.77 & \textcolor{black}{\rule{7mm}{2mm}}  \\
1 &          &   IntTest         & 33 &   42 & 2.54 & \textcolor{black}{\rule{25mm}{2mm}}  \\
1 &          &   SysTest         & 24 &  108 & 1.85 & \textcolor{black}{\rule{18mm}{2mm}}  \\
\hline\multicolumn{7}{c}{~}  \\
1 & Design   &   DesignReview    & 11 &   16 & 1.00 & \textcolor{black}{\rule{10mm}{2mm}}  \\
1 &   &   DesignInspect   & 8 &   12 & 0.73 & \textcolor{black}{\rule{7mm}{2mm}}  \\
1 &   &   CodeReview      & 10 &   18 & 0.91 & \textcolor{black}{\rule{9mm}{2mm}}  \\
1 &   &   CodeInspect     & 9 &  14 & 0.82 & \textcolor{black}{\rule{8mm}{2mm}}  \\
1 &   &   UnitTest        & 11 &   18 & 1.00 & \textcolor{black}{\rule{10mm}{2mm}}  \\
1 &   &   IntTest         & 17 &   31 & 1.55 & \textcolor{black}{\rule{15mm}{2mm}}  \\
1 &   &   SysTest         & 13 &   18 & 1.18 & \textcolor{black}{\rule{12mm}{2mm}}  \\
1 &   &   AcceptTest      & 14 &   19 & 1.27 & \textcolor{black}{\rule{12mm}{2mm}}  \\
\hline\multicolumn{7}{c}{~}  \\
1 &  Code   &   CodeReview    & 10 &  16 & 1.00 & \textcolor{black}{\rule{10mm}{2mm}}  \\
1 &    &   CodeInspect   & 10 &   15 & 1.00 & \textcolor{black}{\rule{10mm}{2mm}}  \\
1 &   &   UnitTest      & 12 &   20 & 1.20 & \textcolor{black}{\rule{12mm}{2mm}}  \\
1 &    &   IntTest       & 14 &   25 & 1.40 & \textcolor{black}{\rule{14mm}{2mm}}  \\
1 &   &   SysTest       & 13 &   20 & 1.30 & \textcolor{black}{\rule{13mm}{2mm}}  \\
1 &    &   AcceptTest    & 16 &   25 & 1.60 & \textcolor{black}{\rule{16mm}{2mm}}  \\

\end{tabular}

\end{center}
\caption{Median times to resolve issues seen in the SEI TSP data. For an explanation of this figure,
see \tion{171projects}.}
\label{fig:raw}
\end{figure}
 \begin{figure}[!t]
\begin{center}
\scriptsize
\begin{tabular}{c|lr|r|rl}
  & \multicolumn{2}{c|}{ } &  Percentiles & \multicolumn{2}{c}{Growth with respect to earliest phase  } \\ 
   & \multicolumn{2}{c|}{Phase} &  (units= miniutes) & \multicolumn{2}{c}{(unitless ratios of two time values) } \\\hline

  rank & injected & removed &   90th   &  \multicolumn{2}{c}{90th percentile growth} \\ 
\hline\multicolumn{6}{c}{~}  \\
1 &  Before   &   DesignInspect     & 32  &  1.00 & \textcolor{black}{\rule{10mm}{2mm}}\\
1 &   &   CodeReview      &   31   &    0.97 & \textcolor{black}{\rule{9mm}{2mm}}\\
1 &   &   CodeInspect     &   32   &    1.00 & \textcolor{black}{\rule{10mm}{2mm}}\\
1 &   &   UnitTest        &   45  &    1.41 & \textcolor{black}{\rule{14mm}{2mm}}\\
1 &   &   IntTest         &   63  &   1.97 & \textcolor{black}{\rule{19mm}{2mm}}\\
1 &   &   SysTest         &   46   &   1.44 & \textcolor{black}{\rule{14mm}{2mm}}\\  
\hline\multicolumn{6}{c}{~}  \\
1 &  Planning     &   ReqtsReview       & 35   &   1.00 & \textcolor{black}{\rule{10mm}{2mm}}\\
1 &               &   DesignInspect     & 31  &   0.89 & \textcolor{black}{\rule{8mm}{2mm}}\\
2 &               &   UnitTest          & 53 &   1.51 & \textcolor{black}{\rule{15mm}{2mm}}\\
\hline\multicolumn{6}{c}{~}  \\
1 &  Reqts   &   ReqtsReview      & 42 &    1.00 & \textcolor{black}{\rule{10mm}{2mm}}\\
1 &          &   ReqtsInspect      & 40 &     0.95 & \textcolor{black}{\rule{9mm}{2mm}}\\
1 &          &   DesignReview      & 34 &    0.81 & \textcolor{black}{\rule{8mm}{2mm}}\\
1 &          &   DesignInspect     & 38 &     0.90 & \textcolor{black}{\rule{9mm}{2mm}}\\
1 &          &   CodeInspect      & 45 &   1.07 & \textcolor{black}{\rule{10mm}{2mm}}\\
1 &          &   UnitTest          & 40 &   0.95 & \textcolor{black}{\rule{9mm}{2mm}}\\
1 &          &   IntTest          & 95 &      2.26 & \textcolor{black}{\rule{22mm}{2mm}}\\
1 &          &   SysTest          & 126 &    3.00 & \textcolor{black}{\rule{30mm}{2mm}}\\
\hline\multicolumn{6}{c}{~}  \\
1 & Design   &   DesignReview     & 37 &   1.00 & \textcolor{black}{\rule{10mm}{2mm}}\\
1 &          &   DesignInspect    & 28 &    0.76 & \textcolor{black}{\rule{7mm}{2mm}}\\
1 &   &   CodeReview        & 40 &     1.08 & \textcolor{black}{\rule{10mm}{2mm}}\\
1 &   &   CodeInspect      & 33 &     0.89 & \textcolor{black}{\rule{8mm}{2mm}}\\
1 &   &   UnitTest          & 41 &     1.11 & \textcolor{black}{\rule{12mm}{2mm}}\\
1 &   &   IntTest          & 75 &    2.03 & \textcolor{black}{\rule{20mm}{2mm}}\\
1 &   &   SysTest           & 40 &     1.08 & \textcolor{black}{\rule{10mm}{2mm}}\\
1 &   &   AcceptTest       & 44 &     1.19 & \textcolor{black}{\rule{12mm}{2mm}}\\
\hline\multicolumn{6}{c}{~}  \\
1 &  Code   &   CodeReview      & 35    & 1.00 & \textcolor{black}{\rule{10mm}{2mm}}\\
1 &    &   CodeInspect    & 32   & 0.91 & \textcolor{black}{\rule{9mm}{2mm}}\\
1 &   &   UnitTest      & 45  & 1.29 & \textcolor{black}{\rule{12mm}{2mm}}\\
1 &    &   IntTest         & 58   & 1.66 & \textcolor{black}{\rule{16mm}{2mm}}\\
1 &   &   SysTest        & 47 &    1.34 & \textcolor{black}{\rule{13mm}{2mm}}\\
1 &    &   AcceptTest      & 60 &   1.71 & \textcolor{black}{\rule{17mm}{2mm}}\\

\end{tabular}
\end{center}
\caption{90th percentile times to resolve issues seen in the SEI TSP data.   Same format as \fig{raw} (but here we look at 90th percentile
outliers while   \fig{raw} explored the central tendencies of the data).}\label{fig:last}
\end{figure}
\subsection{Observations from 171 Projects}\label{sect:171projects}

The count by phase in which defects were removed is shown in Figure~\ref{fig:fix-phase-dist}. Defects are counted only if they they escape the introduction phase unless a bad fix introduces a new defect. These secondary defects occur almost exclusively in test and very rarely in an inspection. 
A high percentage of defects (44\%) were found and fixed in the early phases, i.e., prior to coding. This distribution is similar to that observed for other projects that emphasized investment in software engineering quality assurance practices. For example, Jones and Bonsignour report 52\% of pretest defects removed before entering implementation, for large projects that focus on upfront defect removal techniques \cite{jones12}. NASA robotics projects had a slightly higher percentage (58\%) of defects removed before implementation began, although these had invested in independent verification and validation on top of other forms of defect removal \cite{me08a}.  

\fig{raw} and \fig{last} show the  50th and 90th  percentile (respectively)
of the time spent resolving issues
(note that, in TSP, when developers see issues, they enter {\em review} or 
{\em inspect} or {\em test}
until that issue is retired).
These values include all the time required  to (a)~collect data and realize there is an error;
(b)~prepare a fix;  and (c)~apply some validation
procedure to check the fix (e.g. discuss it with a colleague or execute some tests).

To understand that figure, we offer the following notes:
\bi
\item
Shown here are the 50th/90th percentiles of issue resolution times for issues injected in phase $P_O$ and resolved in phase $P_r$ (these values are calculated
by sorting all resolution time, then reporting the middle values of that sort).
\item
The ``IQR'' column shows the ``inter-quartile range''; i.e. the range of values representing the 75th - 25th percentile range
\item
The results in that figure are  split out according to issues that were fixed in phase $P_r$ after
being introduced in phase $P_0$. The data are sub-divided into tables according to $P_0$;
i.e. according to {\em before, planning, requirements, design} or {\em  code}. 
\item
The left-hand-side ``rank'' column shows the result of the Scott-Knott ranking procedure described in \tion{stats}. These statistical results were applied separately
to each group {\em Before, Planning, Reqts, Design, Code}.
Recall from \S5.5 that if all the fix times within a group
were statistically insignificantly different, then they all earn ``rank=1''.
Note that
most treatments achieved the same ranks i.e. they were found to be insignificantly different  from each other (the one exception is  within the  Planning:UnitTest results 
where UnitTests were ranked 2).
\item
The right-hand-side bars
show the relative sizes of the increases for the  50th (median)   percentile values. These increases are calculated with respect to the first value in each section ``Before, Planning, Reqts. Design, Code''.
\item
These right-hand-side bars are unitless since they are ratios. For example,
on the last line of \fig{raw}, issues injected during coding and fixed in SysTest take 13 minutes (median) to resolve. This is 130\% more than the 10 minutes (median) required
to resolve coding issues during CodeReview. The right-hand-side bar visually represents that 130\%.
\ei
Technical note: to ensure   representativeness, we display examples
where there exist at least $N\ge 30$ examples\footnote{We selected 30
for this threshold via the central limit theorem~\cite{maxwelldata}.} of issues {\em injected} in phase $P_0$ then
removed in phase $P_r$.

The two key features of \fig{raw} and  \fig{last} are:
\be
\item Nowhere in
these results do we see the kind of very large increases reported in the papers
documenting DIE; neither in the median fix times of \fig{raw}
or at the 90th percentile level of \fig{last}. For example, consider the ratio of the issue resolution time
between {\em Before/DesignInspect} and {\em Before/SysTest} result of \fig{raw}. That ratio is 1.11 which is far smaller
than the scale ups seen in \fig{b81}.  
\item
Nearly all the supposed increases seen in \fig{raw} and \fig{last} are insignificantly different
to the other treatments.  The left hand column of \fig{raw} shows the results of the Scott-Knott statistical tests. Note that nearly all
the treatments have the same rank (``1''); i.e. usually there is no statistically significant difference in the 
time to resolve issues. The only exception here is Planning:UnitTest which is ranked ``2'' but even here, the scale up is merely
a factor of 3, and not the exponential increase promised by classic reports of the delayed issue effect.
\ee
One possible explanation for the lack of a DIE effect is that we are looking
broadly at the entire data set but not at specific stratifications. To address that concern,
we spent some time reproducing these figures for various subsets of our data.
That proved to be an unfruitful-- no stratification was found that 
contained an exponential expansion in the time to fix issues. The reason for this was  the small size
of those stratifications  exacerbated the large IQR's seen in this data\footnote{Recall that in a sorted list of numbers,
the inter-quartile range, or IQR, is the difference between the 
  75th and 25th percentile value.}.
Our 171 projects   stratify into subsets of varying sizes. 
The two largest subsets contained only 17 and 12 projects, with numerous much smaller stratifications.
Reasoning over such small samples
is problematic in the general case and, in the case of our data, it is even more problematic due to
the large IQRs of the data.
(To see these large IQRs,  please compare the 50th percentile and IQR columns of Figure~\ref{fig:raw}, where  most of the IQRs are larger than the 50th percentile; i.e. software
data exhibits large variances, which in this case are exacerbated by the smaller samples seen in the stratifications).
Our conclusion from exploring the stratifications is that, given the currently available data, we cannot check
for a DIE effect in subsets of this data.

Before moving on, we comment on some of the counter-intuitive
results in these figures. Consider, for example, the ``Reqts'' results of  \fig{raw} where
the time required to fix issues actually tends to {\em decrease} the longer they are left in the system. In terms of explaining this result, the key thing is the left-hand-side statistical ranking: all these treatments were found to be statistically indistinguishable. In such a set of treatments, the observed
difference may {\em not} be a causal effect; rather, it may just
be the result of  random noise.

\section{Threats to Validity}

Threats to validity are reported according to the four categories described in Wohlin et al.~\cite{wohlin2012}, which are drawn from Cook and Campbell~\cite{cook1979}. 

\subsection{Conclusion Validity}
\label{sect:conclusion_validity}
Threats to conclusion validity are ``issues that affect the ability to draw the correct conclusion about relations between the treatment and the outcome''~\cite{wohlin2012}. We do not have a traditional treatment or control as in a classical experiment. Instead, we evaluate if the DIE holds in a modern data set. The data set is comprised of TSP projects, so the treatment could be misconstrued as TSP, but this is not that case as we do not have an experimental control to compare TSP against. 

\textit{Low statistical power}: Our data set is comprised of 47,376 defect logs. Our primary analysis in \fig{raw} is based on injection-removal phase pairs whose with sample size $>30$. The justification for the statistical techniques used in this paper is provided in \tion{stats}.

\textit{Reliability of measures}: The base measures in this study described in \tion{data-collection} are \textit{defects} recorded in TSP defect logs and \textit{time} reported in time tracking logs. The primary threats to the reliability of these measures are: that the definition of a defect varies between projects and that time is not reported accurately or consistently. The reliability of the time reporting is discussed in \tion{data_integrity}. Time is reported on a level of minutes. We do not have a precise assessment of the error margin for time reporting. Some developers are less precise with time or estimates. Nonetheless, we have applied several tests to verify that the data is accurate. First we compare entries from the defect and time logs to verify that defect log times-to-fix sum to less than the total time log effort in the phase. Second, time log time stamps must be consistent with both the the time stamps and phase for defect in the defect log. Third, we applied a Benford test on the leading digits from the time log  and defect log times to estimate the number data entries that do not result from a natural process (that is, guessed or estimated rather than measured values) ~\cite{shirai14}.  Based on these tests we believe that greater than 90\% of the time log data is recorded in real time. The fidelity and consistency of data will be subject of a future paper. 

We assume that each team has similar defect recording practices, and the TSP coaching provides guidance on what constitutes a defect. Nonetheless, individual developers and teams may apply their own internal rules for filtering defects, which would lead to inconsistent reporting thresholds among the projects in our sample. A related issue is that we assume developers correctly report in which phases a defect was injected and corrected. One point of variation is the measurement framework that identifies process phases and joins the effort to a size measurement framework. Individual projects may choose to implement a different framework, for example adding phases for specific types of development (for example, adding static analysis or special testing or a non-standard size unit).

Certainly, if the defect and time reporting was done
incorrectly in this study, then all our results must be questioned.
However,  this issue threatens {\em every} study on the delayed
issue effect-- so if our results are to be doubted on this score,
then all prior work that reported the delayed issue effect should
also be doubted.  In TSP, developers are trained and supplied with templates for defect and time tracking, all data entry is double-checked by the team TSP coach, and developers are required to analyze their data to make process improvements. That is, TSP developers are always testing
if their project insights are accurate. In such an environment,
it is more likely that they will accurately identify the injection phase.

\textit{Reliability of treatment implementation}: 
Although TSP is not prescriptive about the development process, goals, or strategy, TSP provides precise guidance and training for data gathering. The guidance for logging time and defects is precisely defined. All tasks should be logged as the work is performed with a stopwatch tool. All defects that escape a phase must be logged. All data fields for each defect must be completed.

There are a number of reasons to believe that the data are consistent between developers and between projects. First, developers receive PSP training, during which instructors focus on complete and accurate data gathering. Second, each project that submitted data had a certified TSP coach responsible for evaluating  process adherence and submitting the data. Third, because the teams use their data to manage the projects the team is motivated to collect complete, accurate, and precise data otherwise the data gathering and analysis would be wasted effort. Fourth, process fidelity issues are apparent to the TSP coach as missing or inconsistent data (e.g. time and defect logs do not match, log entries have excessive rounding, or a developer is an outlier). Fifth, 15 of the projects received a TSP Certification in which process fidelity was evaluated independently by an observer and data analyst examining data internal consistency and consistency with  distributional properties known to consistent among all projects and team members. Sixth, all projects in this sample used the same data gathering tool. Nonetheless, some variations exist.


\subsection{Internal validity}
Threats to internal validity concern the causal relationship between the treatment and the outcome~\cite{wohlin2012}. Again, we do not consider TSP as a treatment, but we observe that the DIE does not hold in the TSP data set. Nonetheless, it is useful to consider threats to internal validity at an abstract level between the software engineering milieu that generated the original DIE observations and today's context where TSP was applied.

\textit{History}: Many technological advances have occurred in the time between when DIE was originally observed in the late 70s and today. Processors are more powerful, memory is cheap, programming languages are more expressive, developer tools are more advanced, access to information is easier via the Internet, and significant evolutions in programming paradigms and software process have been realized in the past 40 years. In addition to the risk-oriented, disciplined nature of TSP, any or all of these additional historical factors may have contributed to the lack of the delayed issue effect in our data. 

\textit{Instrumentation}: The forms by which the TSP defect and time data are collected have been studied and matured over 20 years. Conversely, we do not find much documented evidence on how time and defects are reported for the original DIE papers (see \tion{belief}). Thus, we cannot be assured that reporting and data capture were not a significant influence on the delayed issue effect in the original papers.

\textit{Interactions with selection}: As described in \tion{data-collection}, all TSP teams are required to contribute time and defect data to the SEI, and thus there should be no selection bias in this sample compared to the overall population of TSP projects. However, there is likely selection bias in the teams that elect to use TSP compared to the entire population of software development teams. We do not have a basis for comparing TSP teams to those teams in which the DIE was originally observed.

\subsection{Construct validity} \label{sect:construct}
Construct validity concerns ``generalizing the result of the experiment to the concept or theory behind the experiment''~\cite{wohlin2012}. Thus, do the observations in this paper provide evidence on the general delayed issue effect theory?

\textit{Inadequate pre-operational explication of constructs}: As described in \tion{data-collection}, the measures of defect, time, and cost in the original DIE papers are not clearly defined.\footnote{In retrospect, empirical software engineering studies at that time were extremely rare, and guidance for reporting empirical case studies and experiments have improved substantially. One of the seminal books on quasi-experimentation and reporting of validity concerns, Cook and Campbell~\cite{cook1979}, had not been published when most of the DIE papers were written.} Note that in Figure~\ref{fig:cost-to-fix}, the units of ``cost-to-fix'' are not expressed -- in the source references, cost appears as calendar time, effort, and price. In the TSP, a defect is defined as ``any change to a product, after its construction, that is necessary to make the product correc'' and time to correct a defect includes ``the time to analyze, repair, and validate a defect fix.'' Our analysis of DIE focuses on time as a measure of time-as-effort (persons * time). 

The data used in this analysis does not extend into post-delivery deployment. As mentioned in \tion{earlyonset}, every other
paper reporting DIE also reported  early onset of DIE
within the current development. Specifically: those pro-DIE papers reported very large
increases in the time required to resolve issues {\em even before delivery}. That is, extrapolating those
trends it would be possible to predict for a large DIE effect, even before delivering the software.
This is an important point since  Figure~\ref{fig:raw} shows an {\em absence}
of any  large DIE effect during development
(in this data, the greatest increase in difficulty in resolving requirements issues was the 2.16 to 4.37
scale-up seen in the {\em before} to {\em integration testing} 
which is far smaller than the 37 to 250-fold increases reported in \fig{b81} and \fig{cost-to-fix}).

\textit{Mono-method bias}: We only measure the delayed issue effect in terms of defects (as reported by teams) and time (in minutes of effort). To mitigate mono-method bias, additional measures of these constructs would be needed. For example, defects may be segmented into customer-reported defects and pre-release defects. In addition to time-as-effort, calendar time and price to fix (including labor, CPU time, overhead) would provide a more complete picture of the abstraction ``cost to fix a defect''. Further, there are no subjective measures of cost-to-fix, such as the social impact on the team or frustration of the customer.

\textit{Confounding constructs and levels of constructs}: We do not consider the severity of defects in this analysis. Evidence discussed in \cite{Shull02} suggests that low severity defects may exhibit a lower cost to change. Nonetheless, even ``small'' errors have been known to cause enormous damage (e.g., the Mars Climate Orbiter). It is possible that high-severity defects require more effort to fix simply because more people work on them, or conversely, low-severity defects may be fixed quickly simply because they it is easier to do so. High-severity defects are of particular concern in software projects, and even if the number of high-severity defects is low their cost to fix may be extremely large. Note that if such outliers were common in our data, they would appear in the upper percentiles of results. 

\textit{Restricted generalizability across constructs}: While we observe a lack of DIE in the TSP dataset, we examine only the construct of time-to-fix. We do not consider the tradeoffs between time-to-fix and other ''-ilities'', such as maintainability. For example, a low time-to-fix may come at the expense of a more robust solution, i.e., a quick and dirty fix instead of an elegant repair.

\subsection{External Validity}
\label{sect:external_validity}
External validity concerns the generalizability of findings~\cite{wohlin2012} beyond the context of the study. Madigan et al.~\cite{madigan2014} and Carlson \& Morrison~\cite{carlson2009} discuss primarily external validity concerns drawn from studies of large datasets in medicine that are useful for identifying limitations in our study.

\textit{Interaction of selection and treatment}: The most obvious limitation in our study is that the dataset in which we observed no DIE was composed entirely of TSP projects. TSP is a mature process constructed with risk mitigation as its primary purpose. We do not claim that our findings generalize beyond the projects using the TSP process. Similarly, we make no claims regarding generalizability across domains (e.g., defense, banking, games, COTS), scope (\# of features, people, and development length), or organizational features. The purpose of this study is to draw attention to the notion that commonly-held belief of the delayed issue effect may not be a universal truth. This study adds to the evidence offered by the case study in Royce~\cite{Royce98}. Our study invites a further explanation into the causal factors that mitigate DIE.

\textit{Interaction of setting and treatment}: The 171 TSP projects in our data set as well as the case studies in the original DIE papers were all industry projects conducted by software development teams. The TSP projects contain examples of a wide variety
 of systems (ranging from e-commerce web portals to  banking systems) run in a variety of
 ways (agile or  waterfall or some combination of the two). These are realistic settings for contemporary software development teams, though perhaps not representative of all types of projects (see prior paragraph).

\textit{Interaction of history and treatment}: The TSP projects and the original DIE projects took place over several months or years of development. Thus, it is unlikely that the data are substantially influenced by rare events that occurred during project execution.

 \section{Discussion}
 \label{sect:discussion}

Earlier we noted that the delayed issue effect was first reported in 1976 in an era of punch card programming and non-interactive environments~\cite{Boehm76}. We also note that other development practices have changed in ways that could   mitigate the delayed issued effect.  Previously,
most software systems were large, monolithic, and ''write once and maintain
forever.'' Today, even large software systems are trending toward DevOps and cloud-based deployment. Advances in network communications, CPU processing power, memory storage, virtualization, and cloud architectures have enabled faster changes to software, even for large systems. Facebook deploys its 1.5 GB binary blob via BitTorrent in 30 minutes every day~\cite{Facebook}. 
Upgrades to the Microsoft Windows operating system are moving from service patches and major releases to a stream of updates (so there will be no Windows 11- just a stream of continuous updates to what is currently called Windows 10)~\cite{bright15}.


Even organizations that build complex, high assurance systems are turning to agile development processes that purport to address the DIE. For example, agile methods have been advocated for software acquisitions within the US Department of Defense ~\cite{kim13}, and interest and adoption has been growing ~\cite{lapham11}. 
This change in DoD culture is enabled by a separation of baseline architecture
(e.g., the design of an 
aircraft carrier) marked by significant up-front design and the agile development of applications within that architecture.
For the baseline architecture, bad decisions
made early in the life cycle may be too expensive to change and the DIE may still hold.
However, smaller projects within the larger architecture (e.g., lift controls, radar displays) can leverage more agile, interactive development provided that interfaces and architectural requirements are well-defined. 
 
So, is it really surprising that DIE was not observed? Many software engineering technologies have been created precisely to avoid the delayed issue effect by removing risk as early as possible. Boehm's spiral model~\cite{boehm1988spiral}, Humphrey's PSP~\cite{humphrey1995discipline} and TSP~\cite{tsp00}, the Unified Software Development Process~\cite{jacobson1999unified}, and agile methods~\cite{Beck2001a} all in part or in whole focus on removing risk early in the development lifecycle. Indeed, this idea is core to the whole history of iterative and incremental product development dating back to ``plan-do-study-act'' developed at Bell Labs in the 1930's~\cite{larman2003iterative} and popularized by W. Edwards Deming~\cite{deming1986out}. Harter et al. find a statistical correlation between fewer high severity defects and rigorous process discipline in large or complex systems~\cite{harter2012does}. Technical advancements in processing power, storage, networking, and parallelism have combined with a deeper scientific understanding of software construction to enable a whole host of software assurance technologies, from early-phase requirements modeling to automated release testing.

The delayed issue effect may continue to be prevalent in some cases, such as high-assurance software, architecturally complex systems, or in projects with poor engineering discipline. We do not have evidence for or against such claims. However, our data shows that the DIE has been mitigated through some combination of software engineering technology and process in a large set of projects in many domains. Our results are evidence that the software engineering community has been successful in meeting one of its over-arching goals. But our results raise an equally important point - should the DIE persist as a truism (see \S\ref{sect:belief}), or is it a project outcome that can be controlled by software engineering process and technology?

 \section{Conclusion}
 \label{sect:conclusion}
 
In this paper, we explored   the papers and data related to the 
commonly believed {\em delayed issue effect} (that delaying the resolution of issues
very much 
increases the difficulty of completing that  resolution).
Several prominent SE researchers state this effect is a fundamental law of software engineering~\cite{mcconnell01,boehm01,glass02}.
Based on a  survey  of both researchers and practitioners, we  found that
a specific form  of this effect (requirements errors are hardest to fix) is  commonly believed in the community.  

We checked for traces of this effect in 171 projects from the period 2006--2014.
That data held no trace of the delayed issued effect.
To the best of our knowledge, this paper is the  largest study
of this effect yet performed.


We do not claim that this theory {\em never} holds in software projects; just that it cannot be assumed to {\em always} hold, as data have been found that falsify the general theory. Our explanation of the observed lack-of-effect is five-fold. Each of the following explanations is essentially a hypothesis which should be tested against empirical data before we can effectively propose a new theory of the delayed issue effect.
\be
\item The effect might be an historical relic, which does not always hold on contemporary projects. Evidence:
the effect was first described in the era of punch card computing and non-interactive environments.
\item The effect might be intermittent (rather than some fundamental law of software). Evidence: we can  found nearly
as many papers reporting the effect~\cite{Boehm76,Boehm81,steck04,Fagan76,Stephenson76} as otherwise~\cite{Royce98,Boehm80,Shull02}.
\item The effect might be confined to very large systems- in which case it would be
acceptable during development to let smaller to medium
sized projects carry some unresolved issues from early phases into later phases.
\item The effect might be mitigated by modern software development approaches that
encourage change and revision of older parts of the system.
\item The effect might be mitigated by modern software development tools
that simplify the process of large-scale reorganization of software systems.
\ee
  
Our results beg the question: why does the delayed issue effect persist as a truism in software engineering literature? No doubt the original evidence was compelling at the time, but much has changed in the realm of software development in the subsequent 40 years. Possibly the concept of the delayed issue effect (or its more specific description: requirements errors are the hardest to fix)
has persisted because, to use Glass's terms on the subject, it seems to be ``just common sense''\cite{glass02}. 
Nevertheless, in a rapidly changing field such as software engineering, even commonly held rules of thumb must be periodically re-verified. 
Progress in the domain of software analytics has made such periodic checks more cost-effective and feasible, and we argue that an examination of local behaviors (rather than simply accepting global heuristics) can be of significant benefit.

\section*{Acknowledgements}
The authors wish to thank  David Tuma and  Yasutaka Shirai for their work on the SEI databases
that made this analysis possible.
In particular, we thank Tuma Solutions for providing the Team Process Data Warehouse software.
Also, the authors gratefully acknowledge the careful comments of anonymous reviewers from the
FSE and ICSE conferences.
This work was partially funded by an National Science
Foundation grants NSF-CISE 1302169 and CISE 1506586.

This material is based upon work funded and supported by TSP Licensing under Contract No. FA8721-05-C-0003 with Carnegie Mellon University for the operation of the Software Engineering Institute, a federally funded research and development center sponsored by the United States Department of Defense.  This material has been approved for public release and unlimited distribution.  DM-0003956

Personal Software Process$\textsuperscript{SM}$, Team Software Process$\textsuperscript{SM}$, and TSP$\textsuperscript{SM}$ are service marks of Carnegie Mellon University.

\vspace*{0.5mm} 
\balance
\bibliographystyle{plain}
\bibliography{refs} 
\end{document}